\documentclass[epj]{svjour}
\usepackage{graphicx,epsfig,amssymb,amsmath,amstext}
\usepackage{bm}
\usepackage{booktabs}
\usepackage{color}
\usepackage{nicefrac}
\usepackage{dsfont}
\newcommand{\R}{\mathbb{R}} 
 
\newcommand{\C}{\mathbb{C}} 

\newcommand{\Z}{\mathbb{Z}} 
\newcommand{\ii}{\imath}

\newcommand{\dt}[1]{\frac{\mathrm d #1}{\mathrm d t}} % for derivatives
\newcommand{\ket}[1]{\left| #1 \right>} % for Dirac bras
\newcommand{\braket}[2]{\left< #1 |
  #2  \right>} % for Dirac brackets
\begin{document}
\title{Comparing the full time-dependent Bogoliubov--de-Gennes equations to their linear approximation: A numerical investigation}

\author{Christian Hainzl\inst{1} \and Jonathan Seyrich\inst{1}% etc
}
\institute{Mathematisches Institut, Universit\"{a}t T\"{u}bingen,
Auf der Morgenstelle, 72076 T\"{u}bingen, Germany}

\abstract{
 In this paper we report on the results of a numerical study of the nonlinear time-dependent Bardeen--Cooper--Schrieffer (BCS) equations,
 often also denoted as Bogoliubov--de--Gennes (BdG) equations, for
 a one-dimensional system of fermions with contact interaction. We show that, even above the critical temperature, 
 the full equations and their linear approximation give rise to completely
 different evolutions. In contrast to its linearization, the full nonlinear equation does not show any diffusive behavior in the order parameter. 
 This means that the order parameter
 does not follow a Ginzburg--Landau-type of equation, in accordance with a recent theoretical result in \cite{FHSchS}. 
 We include a full description on the numerical implementation of the
 partial differential BCS\slash BdG equations.
 }
%
%\pacs{}
%
%\keywords{}

\maketitle

\section{Introduction}\label{sec:Intro}

When Bardeen, Cooper and Schrieffer, shortly BCS, published one of the most famous papers in physics in 1957 \cite{BCS}, 
giving the first microscopic explanation for superconductivity, a phenomenological theory for the phenomenon had already been
around. Systems close to the critical temperature were described 
with the help of a macroscopic phase-transition parameter introduced by Ginzburg and Landau in 1950 \cite{Ginzburg-Landau}. 
Their theory was the first one to allow for the description of the spatial dependence of the superconductivity inside
superconducting alloys and the first with which to explain type-II superconductors and the hexogonally shaped penetrations by magnetic flux.

As the Ginzburg--Landau theory yields reliable results on the large scale, soon the question arose as to whether this model
can be understood as a macroscopic limit of BCS theory for systems close to the critical temperature. 
Gorkov gave a positive answer to this question for the 
stationary case shortly after the publication of BCS, see~\cite{Gorkov}. 
A rigorous mathematical proof of the convergence was achieved some years ago \cite{proveGLst}.

But what remains unclear and controversial up to this day, in particular in terms of a rigorous derivation,  
is the question whether the time evolution of superconducting systems close to the critical temperature are governed by a Ginzburg--Landau type of equation. 
After first attempts for a derivation of the macroscopic limit had been presented \cite{SteSu,Schmid,AbrTs}, Gorkov and Eliasberg pointed
out that a nonlinear equation could only be valid in a gapless regime \cite{GorEli}.
Still, in \cite{Cyrot,Randeriaetal,Randeria} the authors made a case for a time-dependent Ginzburg--Landau equation for superfluid gases
at temperatures slightly above the critical one. 
The argument is based on the assumption that the nonlinear terms in the BCS\slash BdG equations only lead to small perturbations but do not 
quantitatively change the system's behavior.
In more detail, this would mean that
the projection of the Cooper pair density onto the center of mass direction is governed by a nonlinear dispersive
equation. However, it has been argued recently in \cite{FHSchS} that for a translation invariant homogeneous system close to equilibrium, 
the full BCS\slash BdG equations and their linearization do not yield the same behavior at temperatures close to the critical one. In particular, 
dissipative behavior can only be expected for the linear approximation of the BCS equations but not for the full equations. 

With our work, we demonstrate 
this result by means of a thorough numerical study of the long-term evolution of the BCS equations and their linearization for spatially homogeneous systems 
close to equilibrium 
at temperatures slightly above the critical one. 
For decreasing values of the 
parameter $h$, defined via $T=(1+h^2)T_c$, 
we evolve the full and the linear system over a long time span and track the behavior of the Cooper pair density and the order parameter.
For each values of the small parameter $h$, we find clear differences between the full equation and its linearization. 
Additionally, we see that the full BCS \slash BdG equations yield oscillations in the order parameter about a constant value. Such a behavior has long been predicted
for and already been observed in out-of-equilibrium systems, see, e.g.,~\cite{VolkovKogan,Barankov1,Yuzbashyan}.
Although the focus of our study
is not on oscillations in particular but rather on the long-term behavior of the equations in general, it is interesting that we can 
replicate such oscillations for systems close to equilibrium.

In the realm of numerical analysis, the treatment of quantum
dynamical systems has been of huge interest for many decades (see~\cite{bluebook} for an extensive overview). Various evolution schemes for the linear
Schr\"odinger equation in varying settings have been proposed, see, e.g.,~\cite{FeitFleckSteiger,GrayManolopoulos,BlanesCasasMurua,TalEzerKosloff,ParkLight}. 
Nonlinear Schr\"odinger equations such as the Gross--Pitaevskii equation and 
equations arising from the Hartree and Hartree--Fock approximation of the quantum state have also been devoted attention to, see,
e.g.,~\cite{Bao,Thalhammer,TVZPG,GaucklerLubich}~and~\cite{LubichMCTDH,Lubichvar}. Regarding
the BCS regime, the stationary equations have been treated numerically in \cite{LewinPaul} and the time-dependent BCS\slash BdG equations have been considered
from an analytical perspective in~\cite{HLLSch}.
But, the above-mentioned studies of the out-of-equilibrium dynamics of the BCS equations (\cite{VolkovKogan,Barankov1,Yuzbashyan}) notwithstanding,
to the best
of our knowledge the coupled nonlinear time-dependent BCS equations have not been paid much attention to from a numerical point of view. 
Therefore, we come up with a reliable integration
algorithm for the evolution of the system.

The paper is organized as follows: First, we introduce the system we are considering and the physical background in Section~\ref{sec:BCSeqs}. 
This is followed by a brief 
summary of
the theoretical results of \cite{FHSchS} in Section~\ref{sec:Theor}. Then, we
present our numerical results for the linear and the full equation in Section~\ref{sec:Sim}. Finally, we summarize our main results
in Section~\ref{sec:Con}. A detailed discussion of the initial setup of the system and the numerical implementation is provided in the
Appendix Sections~\ref{sec:tempandgap}~and~\ref{sec:Num}, respectively.

\section{Physical and mathematical background}\label{sec:BCSeqs}
\subsection{Energy functional and BCS equations}
In mathematical terms, BCS theory is a special case of a generalized Hartree--Fock variational principle, itself described by Bogoliubov-theory, 
for the 
density operators $\gamma:\mathcal H\mapsto\mathcal H$ and $\alpha:\mathcal H\mapsto\mathcal H$ acting on the considered Hilbert space $\mathcal H$. 
Those matrices are put together to form the two-by-two operator-valued matrix
\begin{align}\label{eqn-def-Gamma}
 \Gamma:=\begin{pmatrix}\gamma&\alpha\\\overline\alpha&1-\gamma\end{pmatrix},
\end{align} 
see, e.g.,~\cite{BachLiebSolovej} for an introduction.
The entries of the matrix can be represented by means of their momentum distribution $\hat\gamma(k)=\langle  a^\dagger_k a_k \rangle$ and
the pair density $\hat\alpha(k)=\langle a_{k} a_{-k}\rangle$, determining the Cooper pair wave-function via Fourier transform as
$\alpha(x-y) = (2\pi)^{-3/2} \int \hat\alpha(k) e^{i k\cdot (x-y)} d^3k$. 
We suppress spin in our notation; the pair density $\hat\alpha$ is assumed, 
for simplicity, to be a spin singlet. For a one-dimensional translation invariant
system of fermions at temperature
$T$ interacting via a potential $V$, the BCS pressure functional per unit volume is given by
\begin{align}\label{eqn-def-F}
\mathcal F_T(\Gamma)=\int_{\R}(p^2-\mu)\hat\gamma(p)\mathrm dp+\int_{\R} |\alpha(x)|^2V(x)\mathrm dx -TS(\Gamma),
\end{align}
where the entropy $S$ is defined as
\begin{align}
  S(\Gamma):=-\int_{\R}\operatorname{Tr}_{\C^2}\left(\Gamma(p)\log\Gamma(p)\right)\mathrm dp.
\end{align}
The evolutions of 
$\alpha$ and $\gamma$ are given by the time-dependent BCS equations which are also known as Bogoliubov--de--Gennes equations \cite{deGennes}.
In momentum space they can be written conveniently in the self-consistent form
\begin{align}\label{eqn-eom-fundamental}
  \ii\dot\Gamma_t(p)&=[H_{\Gamma_t}(p),\Gamma_t(p)].
\end{align}
Here, the subscript $t$ indicates the time-dependence and the Hamiltonian $H_{\Gamma_t}(p)$ is defined as
\begin{align}
  H_{\Gamma_t}(p)&=\begin{pmatrix}p^2-\mu&2[\hat V\ast\hat\alpha_t](p)\\2[\hat V\ast\overline{\hat\alpha_t}](p)&\mu-p^2\end{pmatrix},
\end{align}
with $\ast$ denoting the convolution. Calculating the upper-left and upper-right entries of the matrix-valued equation~\eqref{eqn-eom-fundamental}, we
arrive at the system of coupled nonlinear equations
\begin{align}
 \ii\dot{\hat\gamma}_t(p)&=2\left[(\hat V\ast\hat\alpha_t)(p)\overline{\hat\alpha_t}(p)-\overline{(\hat V\ast\hat\alpha_t)}(p)\hat\alpha_t(p)\right],
 \label{eqn-dotgammahat-p}\\
 \ii\dot{\hat\alpha}_t(p)&=2(p^2-\mu)\hat\alpha_t(p)+2(\hat V\ast\hat\alpha_t)(p)-4(\hat V\ast\hat\alpha_t)(p)\hat\gamma_t(p).\label{eqn-dotalphahat-p}
\end{align}
\subsection{Contact interactions}
In this paper we concentrate on attractive contact interactions, i.e., potentials of the form 
\begin{align}\label{eqn-def-contact}
V(x)=-a\delta(x), \quad a>0,
\end{align}
which lead to exactly solvable systems in the stationary case.
Not only is
such a potential the most interesting one from a physical model point-of-view but also does it allow us to implement
the terms including a convolution in the equations of motion conveniently as we will illustrate in the numerics Section~\ref{sec:Num}.
\subsection{Initial values}
In this work we consider initial data which, in the stationary case, could be described by the Ginzburg-Landau energy functional for temperatures $T$ close
to the critical temperature $T_c$, i.e., $T=T_c+h^2$ for a small parameter $h\in\R$. For temperatures above $T_c$, the free energy is
minimized by the so-called normal state $\Gamma^\text{N}$ for which $\alpha^\text{N}=0$, $\gamma^\text{N}=\nicefrac1{1+\exp(\nicefrac{(p^2-\mu)}{T})}$. 
For initial data $\Gamma_0$ to be within the range of Ginzburg--Landau, they have to satisfy
\begin{align}\label{eqn-initial-energy-condition}
\mathcal F_T(\Gamma_0)-\mathcal F_T(\Gamma^\text{N})\le \mathcal O(h^4).
\end{align}
This condition can be complied with by choosing 
\begin{align}\label{eqn-Gamma0-def-exp}
\Gamma_0=\frac1{1+e^{\nicefrac{H_{\Delta_0}}{T}}}
\end{align}
with
\begin{align}
  H_{\Delta_0}=\begin{pmatrix}p^2-\mu&-\Delta_0\\-\overline{\Delta_0}&\mu-p^2\end{pmatrix},
\end{align}
where $\Delta_0$ is a small parameter of the order of $h$, see, e.g., \cite{proveGLst}. 
Calculating the right-hand side of the matrix equation~\eqref{eqn-Gamma0-def-exp} gives
\begin{align}\label{eqn-Gamma0-def-detail}
  \Gamma_0=\begin{pmatrix}\hat\gamma_0&\hat\alpha_0\\\overline{\hat\alpha_0}&1-\hat\gamma_0\end{pmatrix}\end{align}
where $\hat\gamma_0$ and $\hat\alpha_0$ take the special form
\begin{align}
  \hat\gamma_0&=\frac12-\frac{p^2-\mu}2\frac{\tanh\left(\frac{\sqrt{(p^2-\mu)^2+|\Delta|^2}}{2T}\right)}{\sqrt{(p^2-\mu)^2+|\Delta|^2}}\\
  \hat\alpha_0&=\frac{\Delta_0}2\frac{\tanh\left(\frac{\sqrt{(p^2-\mu)^2+|\Delta|^2}}{2T}\right)}{\sqrt{(p^2-\mu)^2+|\Delta|^2}}.
\end{align}
In our simulations we choose a temperature which is slightly above the critical temperature for the setting under consideration and set the initial
value for the gap parameter $\Delta_0$ to a non-vanishing value. We explain how to obtain the critical temperature for our setting and how to find physically reasonable
initial values for $\Delta$ in the Appendix Section~\ref{sec:tempandgap}.
\subsection{Ginzburg--Landau and macroscopic parameter}
For the stationary case it is well known that the Ginzburg--Landau theory emerges as the macroscopic limit of the BCS theory. To be more specific, 
define $\ket{\alpha^\ast}$ as the translation invariant minimizer of the BCS functional which, in case of the  contact interaction~\eqref{eqn-def-contact},  can be calculated via
\begin{align}
\hat\alpha^\ast(p)=\frac\Delta2\frac{\tanh\left(\frac{\sqrt{(p^2-\mu)^2+|\Delta|^2}}{2T}\right)}{\sqrt{(p^2-\mu)^2+|\Delta|^2}}.
\end{align}
Then, for the Cooper pair density
$\ket\alpha$ corresponding to the non-translation invariant minimizer of $\mathcal F_T$, the quantity
\begin{align}
  \psi:=\frac1h\braket{\alpha^\ast}{\alpha}
\end{align}
is an approximate solution of the stationary Ginzburg--Landau equation, see, e.g., \cite{FHSS3}. This told,
if there were an analogous relation between the time-dependent BCS and the GL equations,
the order parameter  
\begin{align}\label{eqn-def-psi_t}
  \psi_t:=\frac1h\braket{\alpha^\ast}{\alpha_t}
\end{align}
should, close to $T_c$, approximately satisfy a conventional time-dependent Ginzburg--Landau (TDGL) equation. In the spatially homogeneous case
we are studying in this work, the conventional TDGL equation takes the form 
\begin{align}\label{eqn-def-conventional-TDGL}
  \dot \psi_t=-c_\text{GL,1}\psi_t-c_\text{GL,2}|\psi_t|^2\psi_t,
\end{align}
with some appropriate parameters $c_\text{GL,1}$ and $c_\text{GL,2}$, see, e.g., \cite{Cyrot} and \cite[Eq.~(18)]{Randeria}. 
The parameter $c_\text{GL,1}$ depends on $\nicefrac{(T-T_c)}{(h^2T_c)}$. Crucially, $c_\text{GL,1}$ has the same sign as $(T-T_c)$.
Thus, the TDGL equation is dissipative for temperatures above $T_c$ by definition. 
This implies that if $\psi_t$ could be described by the TDGL for small $h$ it should decay over time. However, we will
demonstrate in Section~\ref{sec:Sim}  that this is not the case, at least for the full non-linear equation.  
The same conclusion has been reached by an analytical investigation recently as we will outline in Section~\ref{sec:Theor}.
\subsection{The linear approximation}
Let us decompose the particle density as 
\begin{align}
  \gamma_t=\gamma_0+\eta_t.
\end{align}
For states  satisfying~\eqref{eqn-initial-energy-condition}
 $\eta_t$ appears to depend quadratically on $\hat \alpha_t$, see, e.g.,~\cite[Eq.~11]{FHSchS}, 
 and it seems legitimate to approximate the full equation by its linearization 
 \begin{align}\label{eqn-dotalphahat-linear}
  \ii\dot{\hat\alpha}_t(p)=2(p^2-\mu)\hat\alpha_t(p)+2(\hat V\ast\hat\alpha_t)(p)(1-2\hat\gamma_0(p)).
\end{align}
However, close to the Fermi-surface the quantity $\eta_t$ is not small 
but the dominant part in the non-linear evolution. Consequently, the full BCS equations~\eqref{eqn-dotgammahat-p}-\eqref{eqn-dotalphahat-p}
and the linearization~\eqref{eqn-dotalphahat-linear} give rise to very different evolutions. Namely, Eq.~\eqref{eqn-dotalphahat-linear} yields a
dissipative behavior in $\psi_t$ whereas the full equations do not as is shown formally in~\cite{FHSchS} and as we confirm by our numerical experiments below.
Let us briefly summarize the results of~\cite{FHSchS} in the next Section.  

\section{Recent mathematical results}\label{sec:Theor}

The BCS time-evolution~\eqref{eqn-eom-fundamental} is studied analytically in \cite{FHSchS}. Based on the work \cite{proveGLst} 
the authors prove in \cite[Theorem 1]{FHSchS} that $|\psi_t|$ does not vanish for any times.
More precisely, it is shown in a very general setting that, if the initial state $\Gamma_0$ is close to the energy of the normal state, i.e.,
$\mathcal F_T(\Gamma_0) - \mathcal F_T(\Gamma^N) \leq O(h^4)$, then 
the corresponding $\psi_t$ satisfies 
\begin{align}\label{eqn-mathematical-bound-nonlinear}
||\psi_t| - |\psi_0|| \leq C h^{1/2},
\end{align}
for an appropriate constant $C$ independent of $h$. 

On the other hand, 
it is shown in \cite{FHSchS} that the solution of the linearized equation~\eqref{eqn-dotalphahat-linear} 
tends to $0$ exponentially fast compared to the system's time scale of $\nicefrac1{h^2}$. In detail, using strategies from perturbation theory, 
it can be derived that 
\begin{align}\label{eqn-mathematical-bound-linear}
|\psi_t|\approx |\psi_0|e^{t\operatorname{Im}\lambda}
\end{align}
holds, where $\lambda$ is a resonance of order $\nicefrac1{h^2}$ which emerges from the 
zero-eigenvalue at $T=T_c$ of the linear operator $\mathcal O=\left(k^2-\mu\right)\tanh^{-1}\left(\frac{k^2-\mu}{2T}\right)+V$. 

The combination of the bounds~\eqref{eqn-mathematical-bound-nonlinear}~and~\eqref{eqn-mathematical-bound-linear} 
shows clearly that the non-dissipative behavior of $\psi_t$ is a purely non-linear effect which takes place solely 
in a tiny neighborhood of the Fermi surface.

Furthermore, using the methods of \cite{FHSchS}, it is straightforward to derive the 
following bound on the derivative 
\begin{align}\label{eqn-psi-dot}
|\dot \psi_t| = \mathcal O \left(\nicefrac1{h}\right).
\end{align}
In other words, although the solution 
$|\psi_t|$ tends to the constant $|\psi_0|$ in the limit $h \to 0$, its derivate might well oscillate more and more --in line with
according predictions for systems which are suddenly perturbed out of equilibrium (\cite{VolkovKogan}). 
These findings are well reproduced in our numerical experiments as we show now.

\section{Simulations}\label{sec:Sim}
In this work we are interested in a qualitative study of the differences between the full BCS\slash BdG equations and their linearization. 
Thus, without loss of generality, we can work in dimensionless units and
set the constant $a$ of the contact interaction and the chemical potential $\mu$ to $a=1$ and $\mu=1$, respectively. 
The initial data for the simulations
are obtained as outlined in the Appendix Section~\ref{sec:tempandgap}. For this, we approximate the integrals in 
Eqs.~\eqref{eqn-Tc-integral}~and~\eqref{eqn-Delta-integral}
by the sum over the discrete momenta we take into account. 
For the sake of reproducibility we add the thus-obtained values for $T_c$ and $\Delta_0$ to the results of our simulations.
For more details on the discretization of the equations under consideration we refer the interested reader to
the Appendix Section~\ref{sec:Num}.
\subsection{Gap as a function of $h$} 
\begin{figure} 
\resizebox{0.45\textwidth}{!}{
  \includegraphics[width=0.4\textwidth]{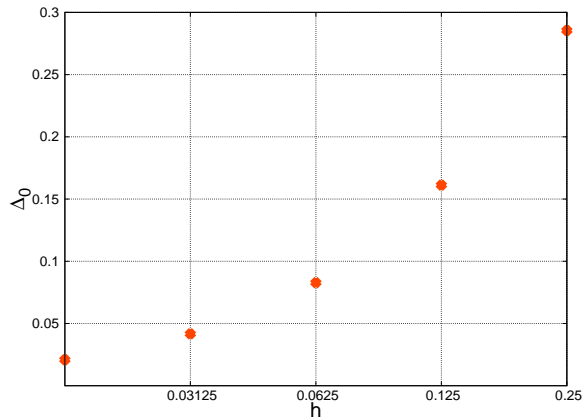}
  }
  \caption{The gap $\Delta_0$ as a function of the semiclassical parameter $h$ in semilogarithmic scale.}
  \label{fig-Delta-of-h}
 \end{figure}

In order not to have to calculate the initial value of the gap parameter which depends on the crucial parameter $h$ 
at the start of each evoluation again, we calculate $\Delta_0$ with the procedure outlined in the Appendix Section~\ref{sec:tempandgap} 
for various $h$ once. 
The interesting result is illustrated in Fig.~\ref{fig-Delta-of-h} where we can see that $\Delta_0$
depends more or less linearly on the crucial parameter.
%We remark that for the critical temperature we get for our system, cf.~Fig.~\ref{fig-Tc-of-M}, the largest possible value of $h$ is $h=\nicefrac14$~\cite{remark2}, since
%for larger $h$ we would have $T=T_c+h^2\gg T_c$. 

Finally, with both $T_c$ and $\Delta_0$ at hand, we are able to present the results of the simulations.
Doing so, we take into account that at temperatures $T=T_c+h^2$, physically interesting dynamics are
expected to occur on a time-scale of $\mathcal O(\nicefrac 1{h^2})$. 
Therefore, we always set $t_\text{end}=\nicefrac1{h^2}$ or $t_\text{end}=\nicefrac2{h^2}$ in the following.
\subsection{Results for $h=\nicefrac14$}
\begin{figure} 
\resizebox{0.45\textwidth}{!}{
  \includegraphics[width=0.4\textwidth]{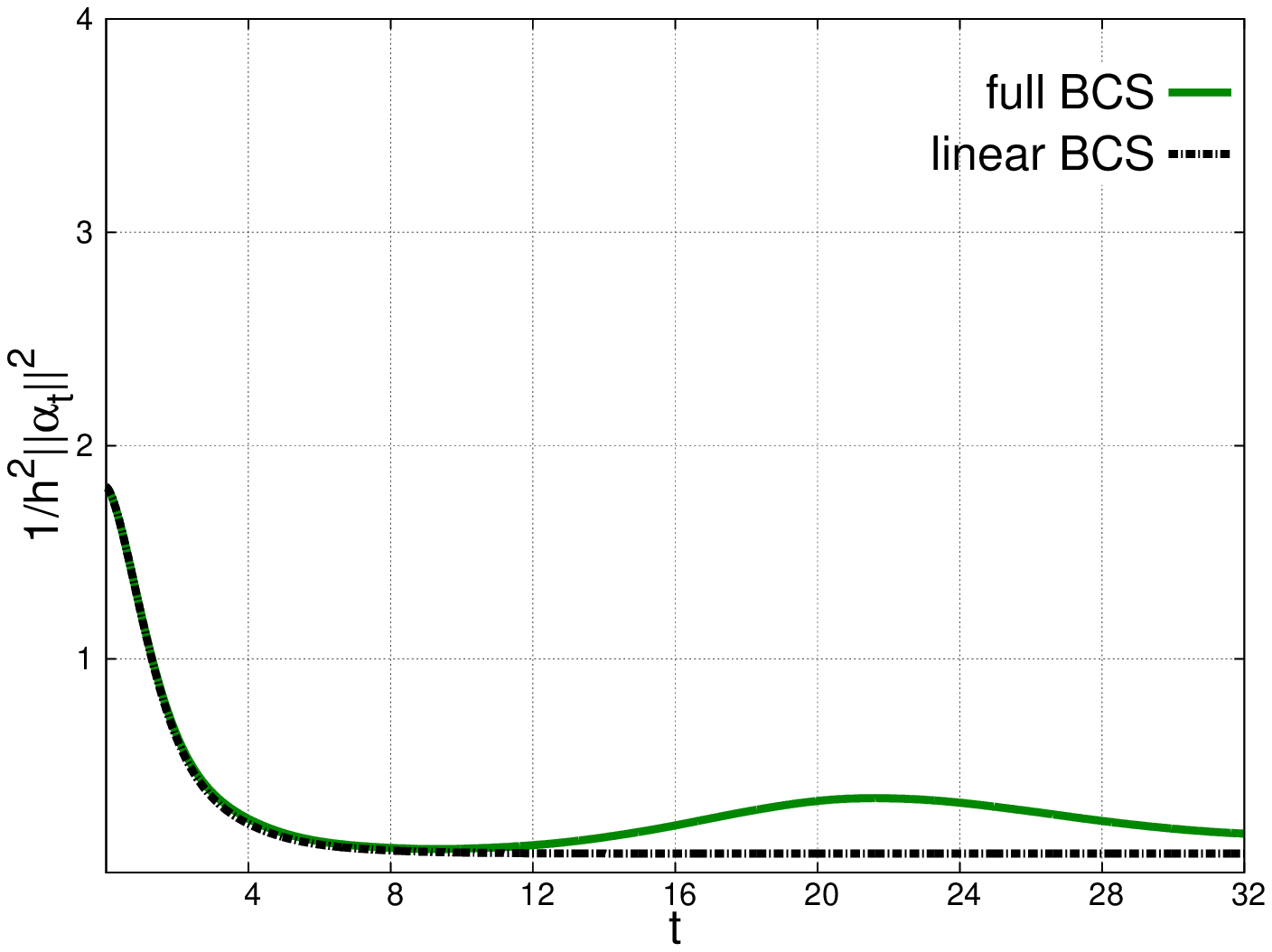}
  }
  \caption{$\nicefrac1{h^2}\|\alpha_t\|^2_2$ as a function of integration time $t$ for $h=\nicefrac14$. The physical parameters are $T_c=0.19$ and $\Delta_0=0.29$.}
  \label{fig-norm-h4}
 \end{figure}

\begin{figure} 
\resizebox{0.45\textwidth}{!}{
  \includegraphics[width=0.4\textwidth]{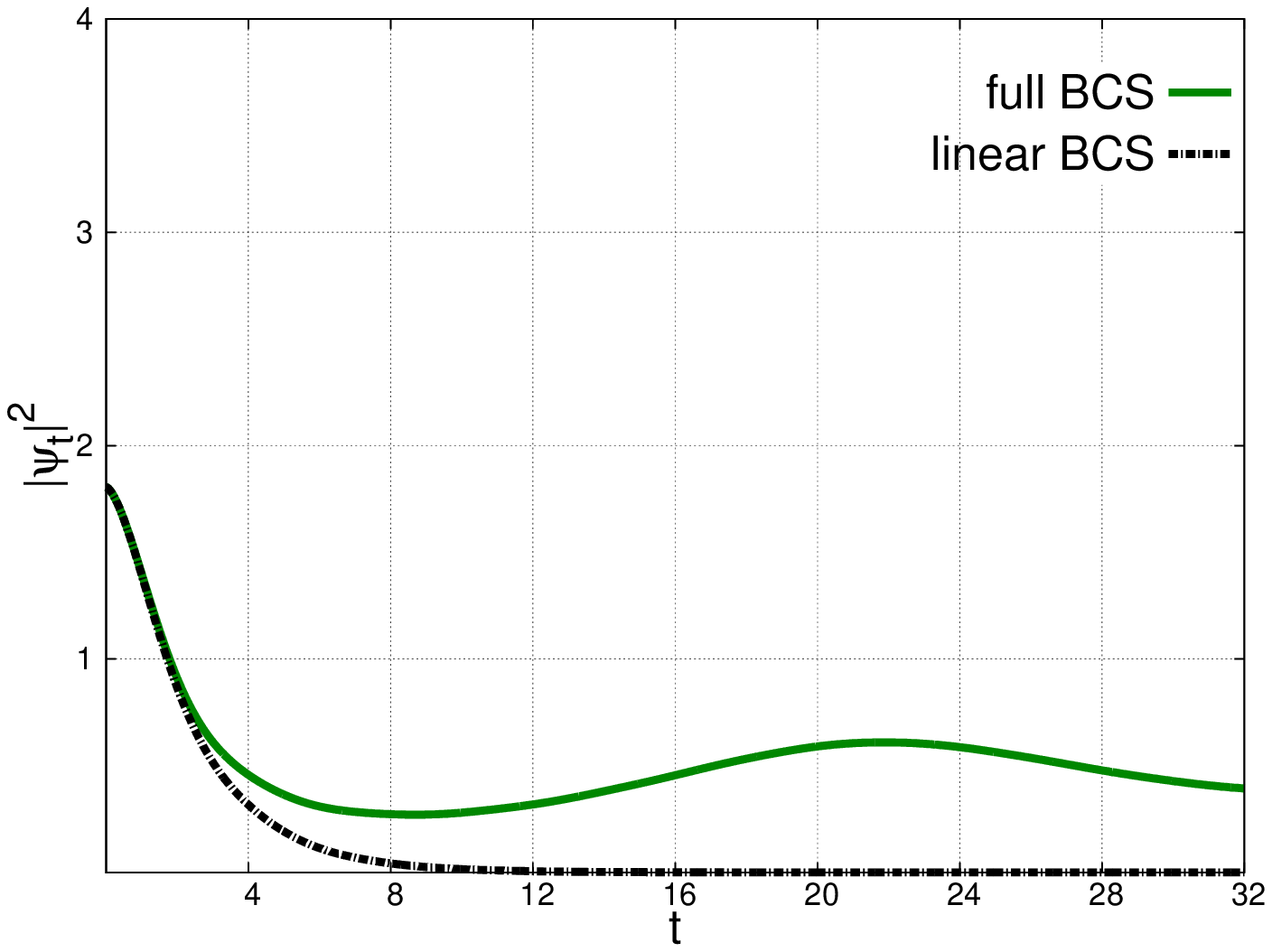}
  }
  \caption{$\psi_t$ as a function of integration time $t$ for $h=\nicefrac14$. The physical parameters are $T_c=0.19$ and $\Delta_0=0.29$.}
  \label{fig-psi-h4}
 \end{figure}

We plot the scaled $L^2$-norm of $\alpha$, which in the discrete setting is given by the sum over the $K$ discrete momenta as
\begin{align}
  \frac1{h^2}\|\alpha_t\|^2_2=\frac1{h^2}\sum_{k=-\nicefrac K2}^{\nicefrac K2-1}|\alpha^K_t(k)|^2,
\end{align}
as well as the modulus of the interesting macroscopic parameter $\psi_t$ introduced in Eq.~\eqref{eqn-def-psi_t}.
We plot the results for both the BCS equation~\eqref{eqn-dotalphahat-p} and its linear approximation~\eqref{eqn-dotalphahat-linear}, 
see Fig.~\ref{fig-norm-h4} and Fig.~\ref{fig-psi-h4}.
 For both quantities, the linear equation leads to exponential decay. 
 The full equation, in contrast, coincides with the linear approximation only for a short period after which both $\|\alpha_t\|$ and $|\psi_t|$ grow again.
\subsection{Results for $h=\nicefrac18$}\label{subsec:h8}
\begin{figure} 
\resizebox{0.45\textwidth}{!}{
  \includegraphics[width=0.4\textwidth]{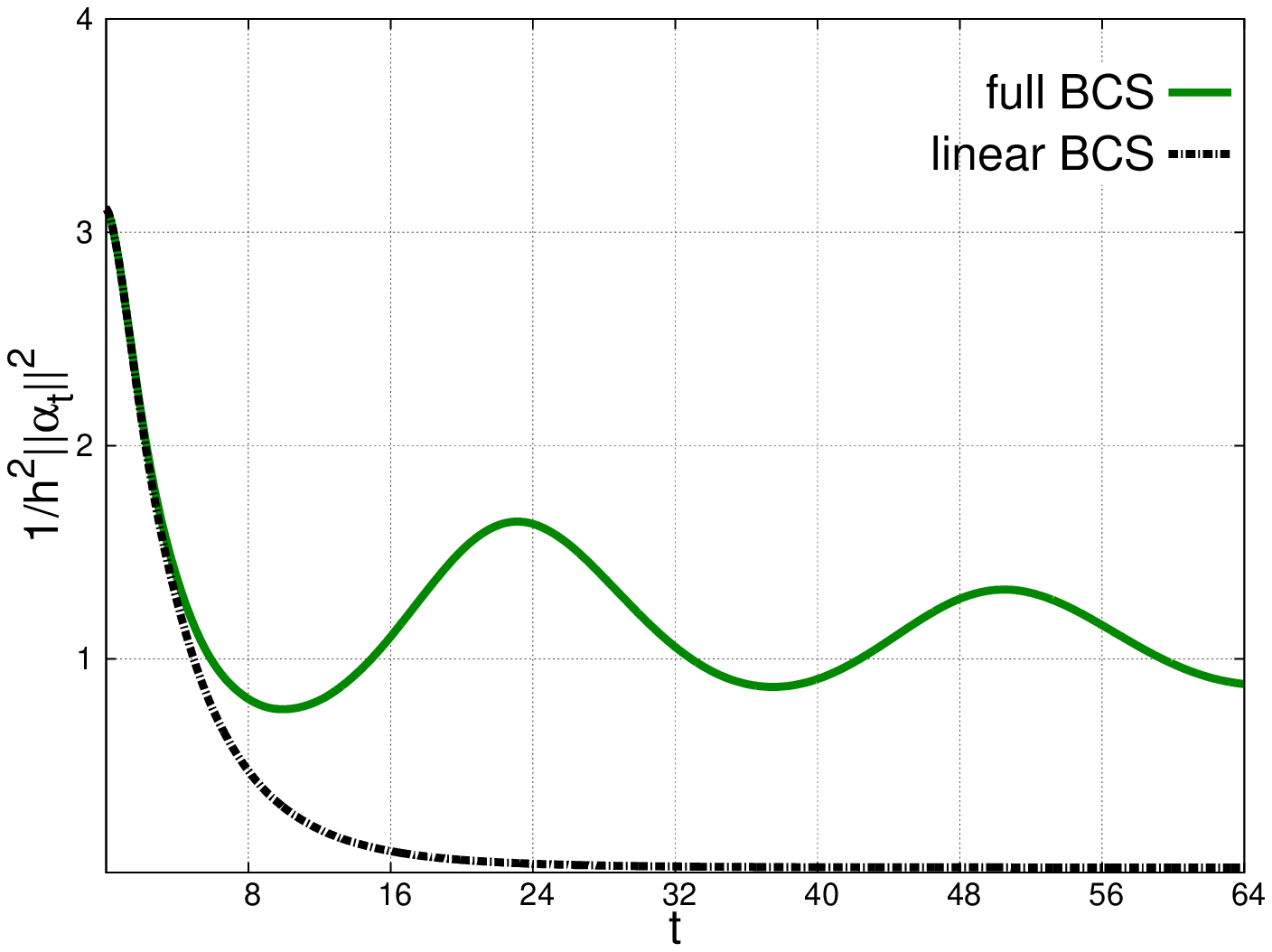}
  }
  \caption{$\nicefrac1{h^2}\|\alpha_t\|^2_2$ as a function of integration time $t$ for $h=\nicefrac18$. The physical parameters are $T_c=0.19$ and $\Delta_0=0.16$.}
  \label{fig-norm-h8}
 \end{figure}

\begin{figure} 
\resizebox{0.45\textwidth}{!}{
  \includegraphics[width=0.4\textwidth]{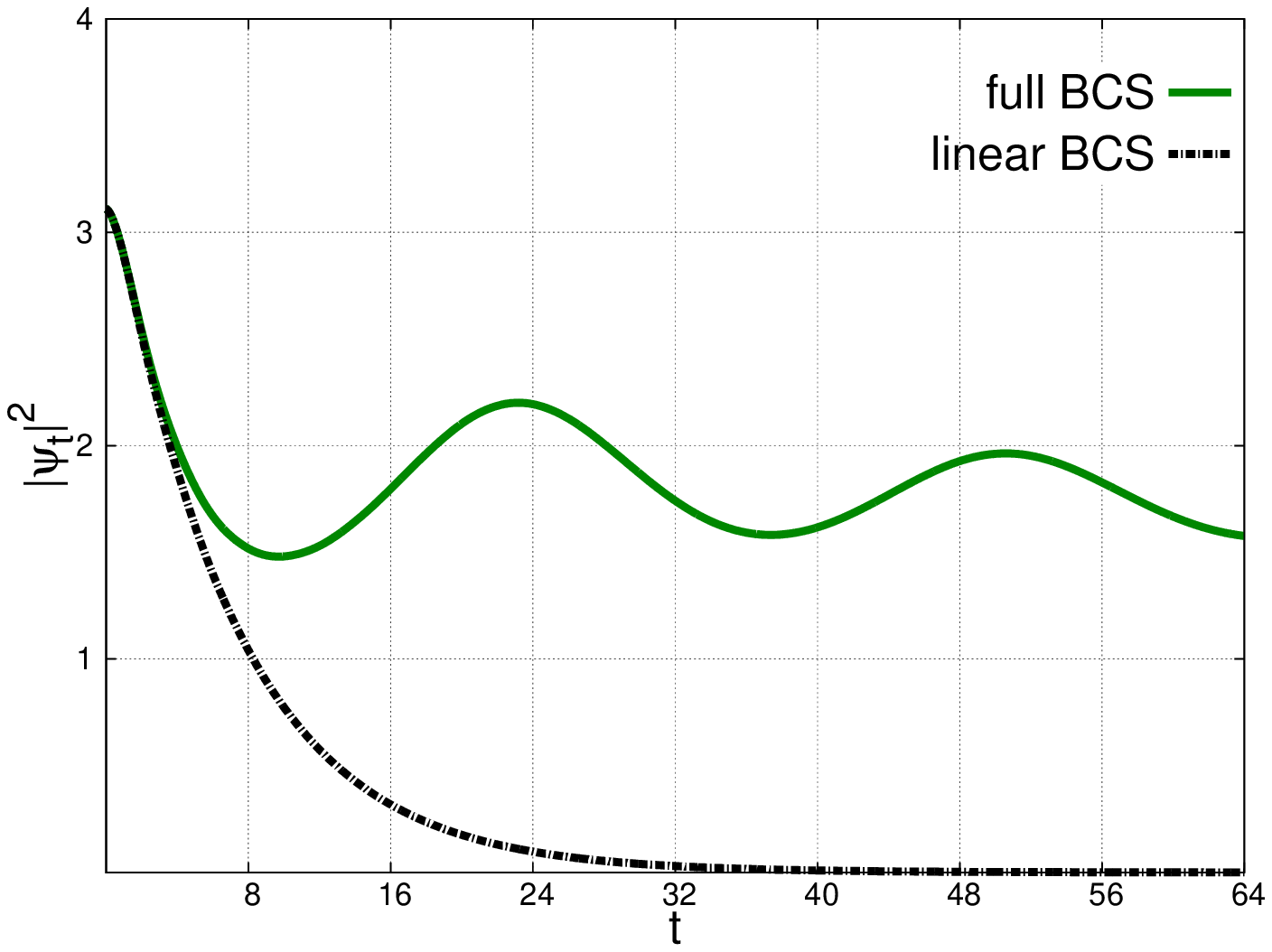}
  }
  \caption{$\psi_t$ as a function of integration time $t$ for $h=\nicefrac18$. The physical parameters are $T_c=0.19$ and $\Delta_0=0.16$.}
  \label{fig-psi-h8}
 \end{figure}

Here, too, we consider the scaled norm of the Cooper pair density and the modulus of the parameter $\psi_t$. 
The results are shown in Fig~\ref{fig-norm-h8} and Fig.~\ref{fig-psi-h8}. 
 Again, the linear evolution equation is clearly diffusive while the full equation yields a similar behavior only for very small times. After a short decline 
 in the beginning
 of the simulation, $\|\alpha_t\|$ and $|\psi_t|$ seem to oscillate. Similar oscillations have
 been predicted by \cite{VolkovKogan} and observed for suddenly perturbed non-equilibrium systems
 in \cite{Yuzbashyan}. Although, as compared to these studies, we work on systems close to equilibrium and slightly above the critical temperature, it 
 is interesting to see that our long-term evolutions show oscilllations which resemble the ones predicted for the out-of-equilibrium case. 
\subsection{Results for $h=\nicefrac1{16}$}
\begin{figure} 
\resizebox{0.45\textwidth}{!}{
  \includegraphics[width=0.4\textwidth]{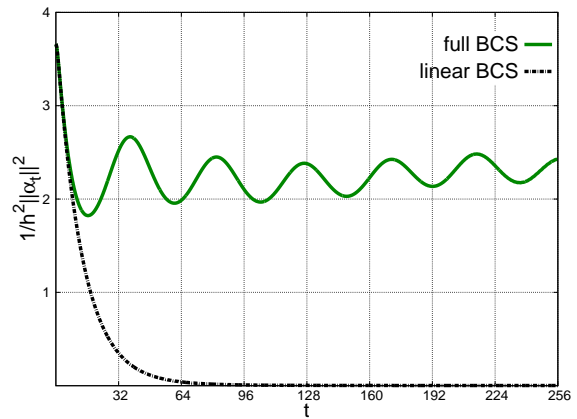}
  }
  \caption{$\nicefrac1{h^2}\|\alpha_t\|^2_2$ as a function of integration time $t$ for $h=\nicefrac1{16}$. The physical parameters are $T_c=0.19$ and $\Delta_0=0.083$.}
  \label{fig-norm-h16}
 \end{figure}

\begin{figure} 
\resizebox{0.45\textwidth}{!}{
  \includegraphics[width=0.4\textwidth]{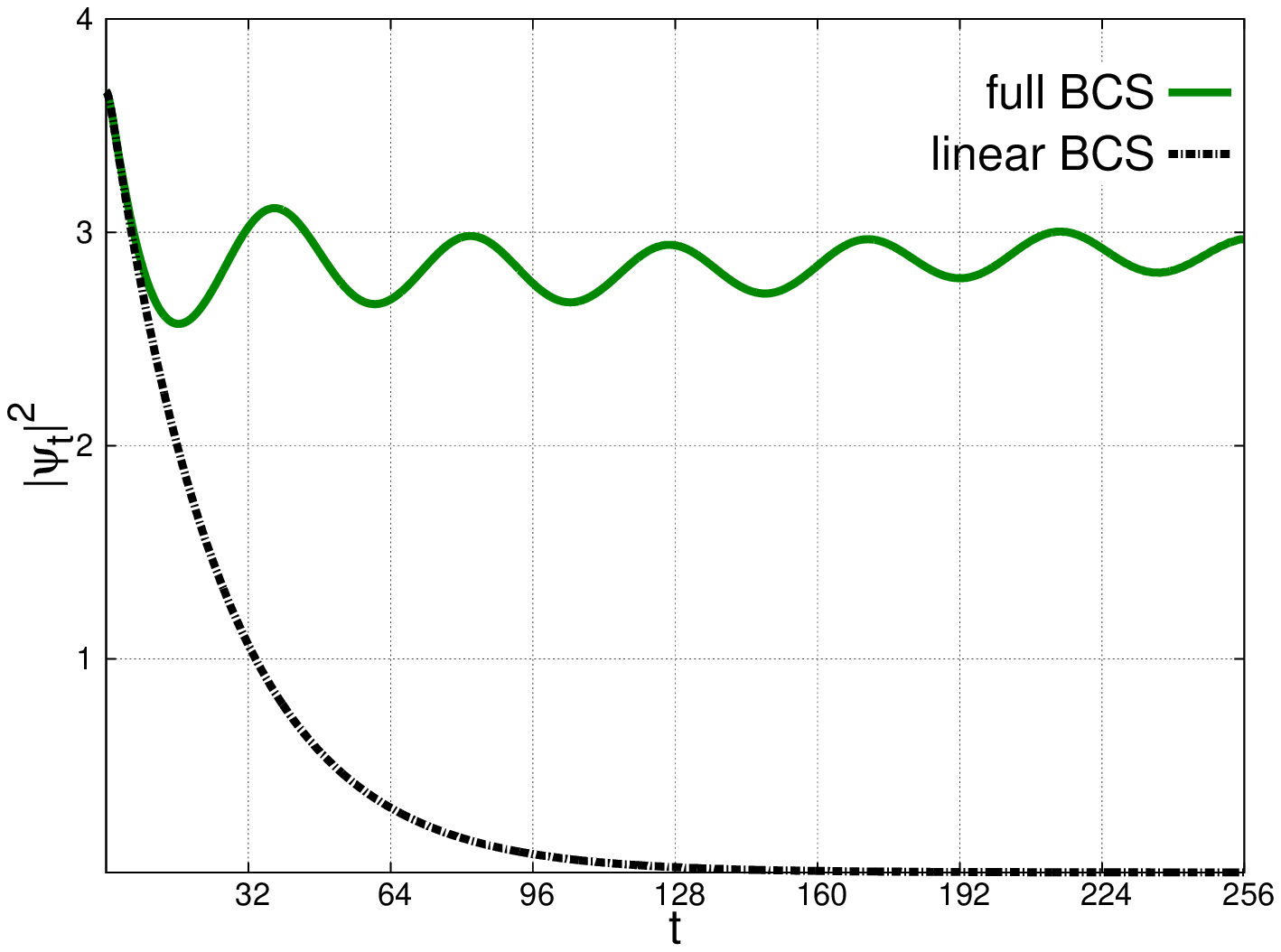}
  }
  \caption{$\psi_t$ as a function of integration time $t$ for $h=\nicefrac1{16}$. The physical parameters are $T_c=0.19$ and $\Delta_0=0.083$.}
  \label{fig-psi-h16}
 \end{figure}

Once more, we depict the time evolution of  $\|\alpha_t\|$ and $|\psi_t|$, cf.~Figs.~\ref{fig-norm-h16}~and~\ref{fig-psi-h16}. 
 The conclusions we can draw from these two plots
are the same as for $h=\nicefrac18$, the only difference being the faster oscilllations in
line with the bound~\eqref{eqn-psi-dot}. Most importantly, even for this small value of $h$, we only observe diffusion for 
the linear approximation which, belying its name, does not approximate the BCS equation for reasonably long time intervals. Let us summarize our results in the concluding Section.

\section{Summary} \label{sec:Con}
We have introduced a reliable integration scheme for the time-dependent BCS equation and its linear approximation in spatially homogeneous settings. 
With the help of these
algorithms, we could perform numerical long-term studies for systems close to equilibrium in order to 
investigate the time-evolution of the order parameter at the limit close to the critical temperature. 
The study shows very clearly that, opposed to the linear case, the full BCS equation does not yield any decay over time in the order parameter $\psi_t$. 
Since the conventional time dependent Ginzburg--Landau equation is dissipative above the critical temperature by definition, it cannot give a valid
macroscopic limit of the full time-dependent BCS\slash BdG equations. It can only be seen as the limit of the linearization of the full equations but
the effects of this linearization could clearly be shown not to be negligible in the considered regime.  
We thus confirm the analysis provided in~\cite{FHSchS}. 

In addition, when evolving the system as described by the non-linear
BCS\slash BdG equations, we observed oscillations in the Cooper pair density and in the order parameter about a finite value which are similar 
to oscillations which have been observed for out-of-equilibrium systems in various works. 
\vspace{\baselineskip}\\
\emph{\textbf{Acknowledgements:}}
 We want to thank Christian Lubich for useful remarks and discussions. This work was partially funded by the DFG grant GRK 1838.
\vspace{\baselineskip}\\
\emph{\textbf{Declaration of contribution}}
 Contribution to the original manuscript was: C.~H. 30\%, J.~S. 70\%.\\ 
 Contribution to the update of the manuscript was: C.~H. 50\%, J.~S. 50\%.

 \appendix
\section{Criticial temperature and initial energy gap}\label{sec:tempandgap}
For translation invariant systems with contact interaction, the cricital temperature $T_c$ is well-known to be given implicitly by
\begin{align}\label{eqn-Tc-integral}
 \frac{2\pi}a=\int_{\R}\frac{\tanh\left(\frac{p^2-\mu}{2T_c}\right)}{p^2-\mu}\mathrm dp,
\end{align}
see, e.g.~\cite{Leggett,NRS,Randeria}. The energy gap $\Delta$ between the superconducting state and the normal state at temperatures beneath the cricital temperature,
in turn, can be obtained from the relation
\begin{align}\label{eqn-Delta-integral}
 \frac{2\pi}a=\int_{\R}\frac{\tanh\left(\frac{\sqrt{(p^2-\mu)^2+|\Delta|^2}}{2T}\right)}{\sqrt{(p^2-\mu)^2+|\Delta|^2}}\mathrm dp.
\end{align}
In order to calculate the critical temperature and a realistic initial value for the gap parameter we thus proceed as follows: 
For a given value of the small crucial parameter $h$, we first determine the critical temperature $T_c$ and set $T=T_c-h^2$. For this
temperature we then search the corresponding gap $\Delta$ following the above definition~\eqref{eqn-Delta-integral} and set $\Delta_0=\Delta$ as its initial state. 
Finally, as we are interested in simulations for temperatures slightly above
the critical temperature, we put $T=T_c+h^2$ and insert this into Eq.~\eqref{eqn-Gamma0-def-detail} together with the just-determined $\Delta_0$. This yields
physically realistic conditions which satisfy the energy constraint~\eqref{eqn-initial-energy-condition}.
\section{Numerical treatment of the equations}\label{sec:Num}
We want to model a system of infinite spacial extension, which, of course, is not possible to achieve on a machine. 
Therefore, we pretend our system to be periodic in space
but with a large enough period. 
\subsection{Finite extension and discrete system}
In the Ginzburg-Landau regime, one often takes into account external potentials that vary on a scale of  $\mathcal O(\nicefrac1h)$ and, consequently,
lead to variations of the system which occur over intervals of that very scale. Thus, a valid model system should have an extension no smaller than those physical variations. But, in 
order to avoid artificial effects due to the periodicity, it is necessary to enlarge this extensions by some multiples of $\nicefrac1h$. For convenience we furthermore include a factor of $2\pi$,
wherefore we consider systems with period $\nicefrac{2\pi N}h$, $1<N\in\mathbb N$. The kernels of the density operators are now functions on $L^2([0,\nicefrac{2\pi N}h]\mapsto\R)$.
In order to simplify the notation, we introduce macroscopic variables via $x_\text{mac}:=\nicefrac hNx$. We end up in a $2\pi$-periodic setting for which the inner product of
two functions $f$ and $g$ is just
\begin{align}
  \braket fg&=\sum_{k\in\Z}\overline{\hat f}\hat g.
\end{align}
The self-consistent BCS equations are now given by
\begin{align}\label{eqn-eom-fundamental-mac}
  \ii \dot\Gamma_t(k)&=[H_{\Gamma_t(K)},\Gamma_t(k)],\quad k\in\Z,
\end{align} 
with the Hamiltonian
\begin{align}
  H_{\Gamma_t}&=\begin{pmatrix}\left(\frac{h^2}{N^2}k^2-\mu\right)&2[\hat V_{Nh}\ast\hat\alpha_t]_k\\2[\hat V_{Nh}\ast\overline{\hat\alpha_t}]_k&\left(\mu-\frac{h^2}{N^2}k^2\right)\end{pmatrix}
\end{align}
and the Fourier transform $\hat V_{Nh}$ of $V_{Nh}(\cdot):=V\left(\nicefrac{N}h\cdot\right)$.
Please note that in the present discrete case the convolution of two summable series $a_k$ and $b_k$ has to be understood as
\begin{align}
 (a\ast b)_k=\sum_{j\in\mathbb Z}a_{k-j}b_j.
\end{align}
\subsection{The equations for a delta potential}
For systems on a large torus with a contact interaction~\eqref{eqn-def-contact}, we can easily see that
\begin{align}
 \hat V_{Nh}\ast\hat\alpha_t=-a\braket\phi\alpha,
\end{align}
where $\phi$ is the state given by $\phi(k)=1$ for all integers $k$. With this, the equations of motion take the convenient form
\begin{align}
 \ii\dot{\hat\gamma}_t(k)&=2a\left[\overline{\braket\phi\alpha}\hat\alpha_t(k)-\braket\phi\alpha\overline{\hat\alpha_t}(k)\right],
 \label{eqn-dotgammahat-p-delta}\\
 \ii\dot{\hat\alpha}_t(k)&=2\left(\frac{h^2}{N^2}k^2-\mu\right)\hat\alpha_t(k)+2a\braket\phi\alpha\left(2\hat\gamma_t(k)-1\right)\label{eqn-dotalphahat-p-delta}
\end{align}
for the nonlinear case and 
\begin{align}\label{eqn-dotalphahat-linear-delta}
  \ii\dot{\hat\alpha}_t(k)&=2\left(\frac{h^2}{N^2}k^2-\mu\right)\hat\alpha_t(k)+2a\braket\phi\alpha\left(2\hat\gamma_0(k)-1\right)
\end{align}
for the linear case. 

Up to now we are still left with an infinite-dimensional system of equations. In order
to solve these numerically, we have to introduce a suitable finite-dimensional subspace. 
\subsection{Space discretization}
As the BCS equations are given in their momentum space representation, it is most convenient to use the so-called \textit{Fourier collocation}.
This means that a $2\pi$-periodic
function $f(x)=\sum_{j\in\mathbb Z}\hat f(j)e^{\ii kx}$ is approximated by
\begin{align}\label{eqn-def-colloc}
  f^K(x)=\sum_{j=-\frac K2}^{\frac K2-1}\hat f^K(j)e^{\ii kx},
\end{align}
where the coefficients $\hat f^K(j)$ are obtained by the discrete Fourier transform of the values $f_j=f\left(\nicefrac{2\pi}Kj\right)$, $j=-\nicefrac K2,...,\nicefrac K2-1$. 
Mathematically speaking we work on the subspace spanned by the first $K$ eigenfunctions of the Laplacian on $[0,2\pi]$. As a consequence, the evolution of the system is given
by the $K$-dimensional system of ordinary differential equations (ODE)
\begin{align}
 &\ii\dot{\hat\gamma}^K_t(k)=2a\left[\overline{\braket{\phi^K}{\alpha^K}}\hat\alpha^K_t(k)-\braket{\phi^K}{\alpha^K}\overline{\hat\alpha^K_t}(k)\right],
 \label{eqn-dotgammahat-k-delta}\\
 &\quad\quad -\frac K2\le k\le \frac K2-1,\nonumber\\
 &\ii\dot{\hat\alpha}^K_t(k)=2\left(\frac{h^2}{N^2}k^2-\mu\right)\hat\alpha^K_t(k)\nonumber\\
 &\phantom{\ii\dot{\hat\alpha}^K_t(k)=}+2a\braket{\phi^K}{\alpha^K}\left(2\hat\gamma^K_t(k)-1\right),\label{eqn-dotalphahat-k-delta}\\
 &\quad\quad -\frac K2\le k\le \frac K2-1,\nonumber
\end{align}
and accordingly for the linear case. From numerical analysis it is well known that~\eqref{eqn-def-colloc} yields a very good approximation to
$2\pi$-periodic functions with the discretization error decreasing rapidly as a function of $K$, see, e.g.~\cite{bluebook}, Chapter III.1.3.

For practical reasons we set $K$ to be an integer power of $2$ so that for a given $\hat\alpha^K_t(j)$, $j=-\nicefrac K2,...,\nicefrac K2-1$,
the corresponding distribution $\alpha^K_t(x)$ at the discrete points $x_j=\frac{2\pi}Kj$ can be computed efficiently with the well-known fast Fourier transform (FFT). 
As we want to resolve phenomena happening on the microscopic scale $\mathcal O(\nicefrac hN)$, we choose 
\begin{align}\label{eqn-K}
  K=M\frac Nh
\end{align}
for a large enough integer $M$. Let us now explain how we solve the system of ODE~\eqref{eqn-dotgammahat-k-delta}--\eqref{eqn-dotalphahat-k-delta}.
\subsection{Solving the system of ordinary differential equations}
We first notice that the Hamiltonian $H_{\Gamma_t}$ is self-adjoint. Thus, the time-evolution of $\Gamma_t$ is a unitary transformation and, hence, its eigenvalues
are preserved. With regard to definition~\eqref{eqn-def-Gamma}, the eigenvalues can be readily computed as
\begin{align}
 \lambda_{1,2}(p)&=\frac12\pm\sqrt{\left(\hat\gamma_t(p)-\frac12\right)^2+|\hat\alpha_t(p)|^2},
\end{align}
and we see that the equality
\begin{align}
 \left(\hat\gamma_t(p)-\frac12\right)^2+|\hat\alpha_t(p)|^2&=\left(\hat\gamma_0(p)-\frac12\right)^2+|\hat\alpha_0(p)|^2
\end{align}
holds. Solving this for $\hat\gamma_t$, we get
\begin{align}\label{eqn-gamma-of-alpha}
  \hat\gamma_t(p)&=\begin{cases}\frac12+\sqrt{h(p)-|\hat\alpha_t(p)|^2}\text{ for $p^2<\mu$,}
  \\\frac12-\sqrt{h(p)-|\hat\alpha_t(p)|^2}\text{ for $p^2\ge\mu$,}\end{cases}
\end{align}
where we have defined the auxiliary function
\begin{align}
  h(p)&:=\left(\hat\gamma_0(p)-\frac12\right)^2+|\hat\alpha_0(p)|^2.
\end{align}
The signs in Eq.~\eqref{eqn-gamma-of-alpha} can be inferred from the initial values we use in this work, cf.~\eqref{eqn-Gamma0-def-detail}. 
They are such that $\hat\gamma_0(p)$ is
greater than $\nicefrac12$ for $\mu>p^2$ and less than or equal to $\nicefrac12$ for $\mu\le p^2$.

Inserting the discrete analogon of Eq.~\eqref{eqn-gamma-of-alpha} into the relevant equation of motion~\eqref{eqn-dotalphahat-k-delta}, 
we get the nonlinear coupled system
of equations
\begin{align}\label{eqn-dotalphahat-reduced}
 \ii\dot{\hat\alpha}^K_t(k)&=2\left(\frac{h^2}{N^2}k^2-\mu\right)\hat\alpha^K_t(k)\nonumber\\
 &\phantom{=}\pm 4a\braket{\phi^K}{\alpha^K}\sqrt{h(k)-|\hat\alpha^K_t(k)|^2},-\frac K2\le k\le \frac K2-1.
\end{align}
This said, we now present our time integration algorithm.
\subsection{Time discretization}
Putting it in a formal way, the system we have to integrate is given by
\begin{align}\label{eqn-def-initial-value-problem}
 \begin{cases}\dt{\mathbf y(t)}&=f(\mathbf y(t)),\\
         \mathbf y(0)&=\mathbf y_0,\end{cases}
\end{align}
with 
\begin{align}
  \mathbf y=\begin{pmatrix}\hat\alpha^K\left(-\nicefrac K2\right)&\hdots&\hat\alpha^K\left(\nicefrac K2-1\right)\end{pmatrix}^T\in \mathbb R^K.
\end{align}
The right hand side of our initial problem can be written as the sum of two terms,
\begin{align}\label{eqn-def-split}
f(\mathbf y)=f_1(\mathbf y)+f_2(\mathbf y),
\end{align}
where $f_1$ represents the linear part which resembles the kinetic part in the Schr\"odinger-equation and $f_2$ is the nonlinear part.
Let $\tau$ denote a time step and $\Phi_{\tau,f}$ the smooth map between $\mathbf y(0)$ and $\mathbf y(\tau)$. Given the special
form~\eqref{eqn-def-split} of the differential equation, one can approximate $\Phi_{\tau,f}$ numerically by
\begin{align}
 \Phi^\text{num}_{\tau,f}(\mathbf y_0)=\left(\Phi_{\nicefrac\tau2,f_1}\circ\Phi_{\tau,f_2}\circ\Phi_{\nicefrac\tau2,f_1}\right)(\mathbf y_0).
\end{align}
This is the well-known \textit{Strang splitting}. Applying it successively yields an approximation to the exact solution at times $t=n\tau$, $n=1,2,...$, 
the error of which decreases quadratically as a function of the step size $\tau$, see, e.g.~\cite{hairerlubichwanner}, Chapter II.5.

The advantage of the Strang splitting is that $\Phi_{\tau,h}$ can be calculated exactly as
\begin{align}
 \Phi_{\tau,f_1}(\cdot)=e^{-\ii2\left(\frac{h^2}{N^2}k^2-\mu\right)\tau}\cdot.
\end{align}
As for $\Phi_{\tau,f_2}$, it has to be approximated due to the nonlinearity. 
For this, we choose a simple Runge-Kutta scheme as proposed by~\cite{nr} whose numerical error is small compared to the error expacted
from the splitting~\cite{remark}. Before starting the simulations, we still need to fix the mentioned discretization parameters $\tau$
and $K$. In our case, $K$ itself
depends on three parameters, cf. Eq.~\eqref{eqn-K}. As $h$ is the semiclassical parameter we want to vary throughout the study, we
have to choose reasonable values for the remaining quantities $M$, $N$ and $\tau$. We first consider $\tau$.
\subsection{Fixing the time discretization parameter}
\begin{figure} 
\resizebox{0.45\textwidth}{!}{
  \includegraphics[width=0.4\textwidth]{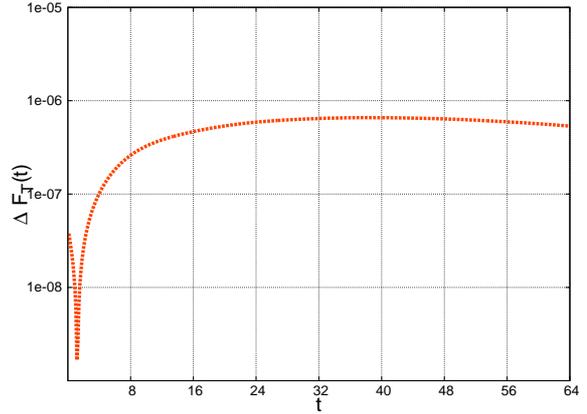}
  }
  \caption{Relative error $\Delta F_T$ of the discretized free energy against integration time $t$ in semilogarithmic scale for $h=\nicefrac18$. 
  The physical parameters are $T_c=0.19$ and $\Delta_0=0.16$.}
  \label{fig-DeltaF}
 \end{figure}

The step size has to be chosen small enough for both the numerical approximation of $\Phi_{\tau,f_2}$ and the Strang splitting
to give accurate results. For our simulations it turned out that reliable results can only be expacted for a step size inversely proportional to $K$.
Playing safe we include a small factor and set $\tau=\nicefrac{0.1}{K}$. As a measure for the time integrator's accuracy, we consider
the discrete analogon of the free energy introduced in Eq.~\eqref{eqn-def-F} above, which is given by
\begin{align}
  F_T^K(\Gamma^K)&=\sum_{k=-\nicefrac K2}^{\nicefrac K2-1}\left(\frac{h^2}{N^2}k^2-\mu\right)\hat\gamma^K(k)\nonumber\\
  &\phantom{=}+
  \frac1{2\pi}\int_0^{2\pi}V(x_\text{mac})|\alpha^K(x_\text{mac})|^2\mathrm dx_\text{mac}-TS(\Gamma^K).
\end{align}
A short calculation yields that this quantity is conserved under the exact flow of the corresponding initial value problem. Therefore, the reliability
of a numerical integration scheme can be checked by tracking the relative error $\Delta F_T$, defined by
\begin{align}
  \Delta F_T(t)=\left|\frac{F_T(\Gamma^K_t)-F_T(\Gamma^K_0)}{F_T(\Gamma^K_0)}\right|,
\end{align}
along the numerical evolution. Recurring to a constant of motion as a criterion of accurateness is a much applied procedure 
in various computational fields, see, e.g.~\cite{Seyrich,Lukes}. Following this line of reasoning, we have verified the accuracy of our time integrator
for every simulation presented below. As an example, we show the plot of $\Delta F_T$ corresponding to the simulations of Subsection~\ref{subsec:h8}
in Fig.~\ref{fig-DeltaF}.
\subsection{Fixing the space discretization parameters}
\begin{figure} 
\resizebox{0.45\textwidth}{!}{
  \includegraphics[width=0.4\textwidth]{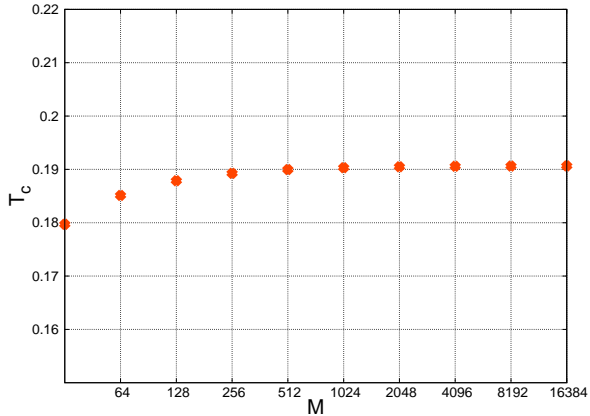}
  }
  \caption{Critical temperature $T_c$ as a function of the number of momenta per unit volume $M$ for $N=8$ and $h=\nicefrac14$.}
  \label{fig-Tc-of-M}
 \end{figure}

We have seen in the previous Subsection that the time step has to be inversely proportional to the dimension of the subspace we are approximating 
our system on. Furthermore, every time step requires a computational effort which grows linearly with $K$. Consequently, the complete CPU time
for a simulation over a given time interval $[0,t_\text{end}]$ is quadratic in $K$. So the dimension of the subspace and, thus, the related $N$ and $M$ 
should be the smallest possible. In order to check how small a $M$ we can choose without any significant loss of accuracy, we fix $N=8$ and $h=\nicefrac14$ and 
calculate the cricital temperature via the discretized version of Eq.~\eqref{eqn-Tc-integral},
\begin{align}
 \frac{2\pi}a=\sum_{k=-\nicefrac K2}^{\nicefrac K2-1}\frac{\tanh\left(\frac{\frac{h^2}{N^2}k^2-\mu}{2T_c}\right)}{\frac{h^2}{N^2}k^2-\mu},
\end{align}
for different values of $M$. The result can be seen in Fig.~\ref{fig-Tc-of-M}.
For different values of $N$ and $h$ we get the same plot. We see that for $M=256$ the critical temperature is still slightly too small. However,
when comparing the evolutions obtained with $M=256$ to the according ones
for $M=512$, the relevant figures are indistinguighable from each another. For the 
sake of efficiency, we thus fix $M=256$ for the rest of this work.

\begin{figure} 
\resizebox{0.45\textwidth}{!}{
  \includegraphics[width=0.4\textwidth]{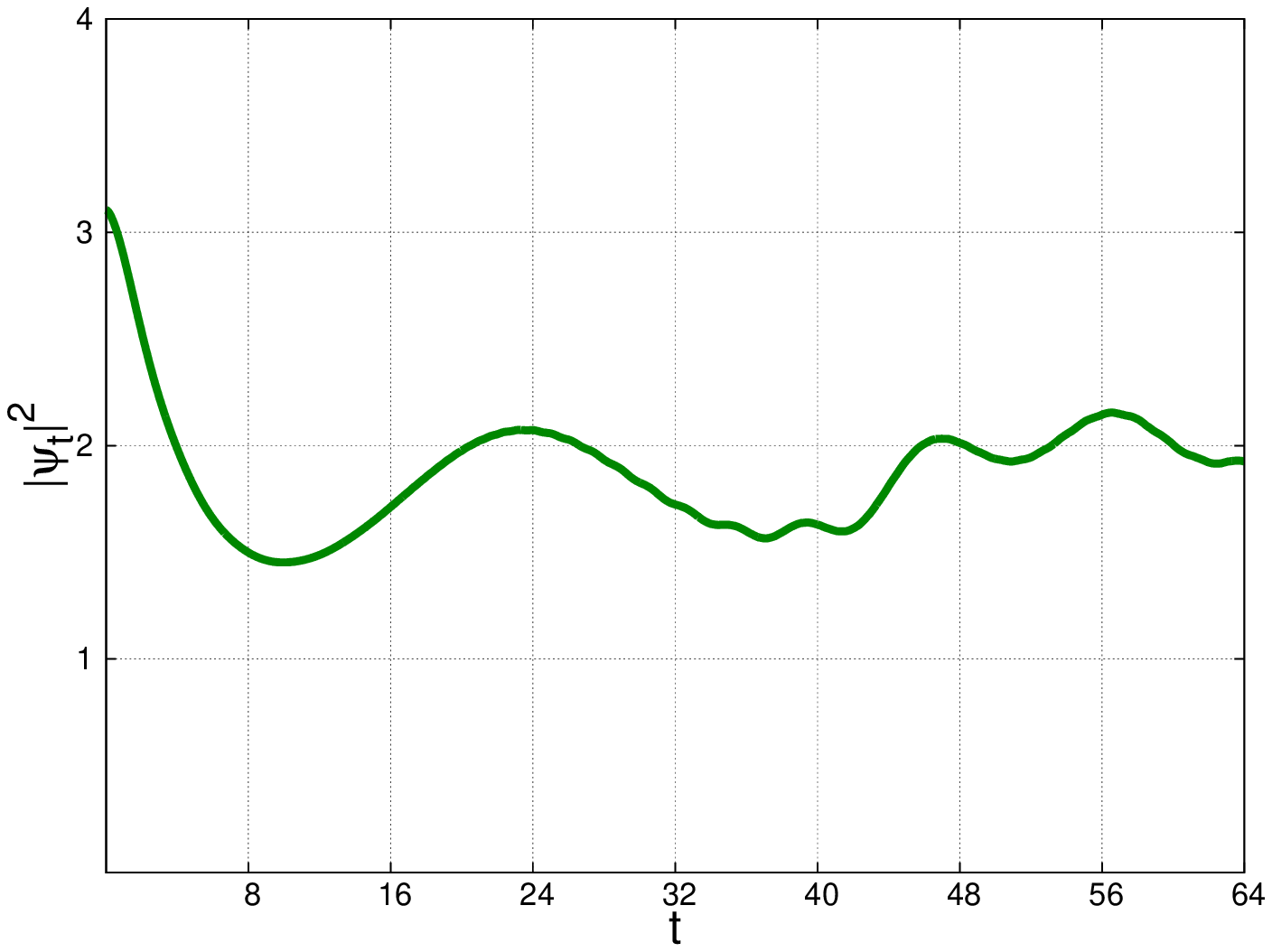}
  }
  \caption{$\psi_t$ as a function of integration time $t$ for $N=4$ and $h=\nicefrac18$. The physical parameters are $T_c=0.19$ and $\Delta_0=0.16$.}
  \label{fig-interference}
 \end{figure}

As for the extension of our interval, $N$, we have to choose it large enough so that the solution cannot reach the boundaries during the simulation. As,
by construction, we work with a periodic setting, a solution reaching one end of the interval would enter again at the other end, thus leading to
unphysical interference. As an example of this numerical artifact, we consider the case $h=\nicefrac18$, $N=4$ and plot the modulus of
$\psi_t$ in Fig.~\ref{fig-interference}.
We observe oscillations for larger $t$ which should not show up in reality, cf.~Subsection~\ref{subsec:h8}. Whenever we encountered such an artifact, we
successively increased $N$ by factors of $2$ until the artifact vanished.


\begin{thebibliography}{}
\bibitem{remark}
  Note that in the linear case $\Phi_{\tau,f_2}$ can also be calculated exactly.
\end{thebibliography}
\end{document}